%% file: main.tex
\documentclass{PoS}
\usepackage{amsfonts} 
\usepackage{amssymb} 
\usepackage{amsmath} 
\usepackage{graphicx} 
\usepackage{subfigure}
\usepackage{array} 
\usepackage{bm} 
\usepackage{longtable} 
\usepackage[dvipsnames]{xcolor} 
\usepackage{comment}
\usepackage{verbatim}
\usepackage{multirow}
\graphicspath{{./figs/}}

\allowdisplaybreaks

\input{macro.tex}

\title{Leptonic decays of $B_{(s)}$ and $D_{(s)}$ using the OK action}

\ShortTitle{$f_{B_s}/f_B$ and $f_{D_s}/f_D$ with the OK action}

\author{Sungwoo Park, Tanmoy Bhattacharya, Rajan Gupta \\
  Theoretical Division T-2, Los Alamos National Laboratory, Los
  Alamos, NM 87545, USA \\
  E-mail: \email{sungwoo@lanl.gov}, \email{tanmoy@lanl.gov},
  \email{rg@lanl.gov}}

\author{Yong-Chull Jang \\
  Physics Department, Brookhaven National Laboratory, Upton, NY 11973,
  USA \\
  E-mail: \email{ypj@bnl.gov}}

\author{\speaker{Benjamin J.~Choi}, Seungyeob Jwa, Sunkyu Lee,
  Weonjong Lee \\
  Lattice Gauge Theory Research Center, CTP, and FPRD,\\
  Department of Physics and Astronomy, Seoul National University,
  Seoul 08826, South Korea \\
  E-mail: \email{wlee@snu.ac.kr}}

\author{Jaehoon Leem\\
  School of Physics, Korea Institute for Advanced Study (KIAS), Seoul
  02455, South Korea \\
  E-mail: \email{leemjaehoon@kias.re.kr}}

\author{LANL-SWME Collaboration}

\abstract{We present recent progress in the lattice calculation of
  leptonic decay constants for $B_{(s)}$ and $D_{(s)}$ mesons using
  the Oktay-Kronfeld (OK) action for charm and bottom valence quarks, whose 
  masses are tuned  non-perturbatively.
  The calculations are done on 6 HISQ ensembles generated by the MILC
  collaboration with $N_f=2+1+1$ flavors.
  We also use the HISQ action for the light spectator quarks. 
  Results are presented for the ratios $f_{B_s}/f_B$ and
  $f_{D_s}/f_D$, which reflect $SU(3)$ flavor symmetry breaking,
  and are independent of the renormalization constants of the axial
  currents.
}

\FullConference{37th International Symposium on Lattice Field Theory -
  Lattice2019\\
  16-22 June 2019\\
  Wuhan, China}

\begin{document}

\section{Introduction}
\label{sec:intr}

The decay constant $f_P$ of a pseudoscalar meson $P$ is defined
by
\begin{align}
  \langle 0 | A^\mu | P \rangle = i p^\mu f_{P} \,, 
  \label{eq:f}
\end{align}
where the external state $|P\rangle$ carries momentum $p^\mu$, and the
axial current $A^\mu = \bar{\psi}_{h} \gamma^\mu \gamma_5 \psi_{l}$.
Here, the subscript $h$ ($l$) represents heavy (light) flavors in
the $B_{(s)}$ and $D_{(s)}$ states.
We use the Oktay-Kronfeld (OK) action \cite{Oktay:2008ex} for valence
heavy quarks $\psi_h$ with $h=b,c$, and the HISQ action
\cite{Follana:2006rc} for valence light quarks $\chi$ that are
  recast into the naive quark field $\psi_l$.
The calculations are done on MILC HISQ ensembles with $N_f=2+1+1$
\cite{Bazavov:2012xda}, whose parameters are summarized in Table \ref{tab:ensembles}. 

\begin{table}[!b]
  \begin{center}
    \begin{tabular}{ l || l | l | c | l l l }
      \hline\hline
      ensemble ID & $a$ (fm) & $N_s^3 \times N_t$ & $M_\pi$
      (MeV) & $am_l$ & $am_s$ & $am_c$ \\
      \hline
      a12m310 & 0.1207(11) & $24^3\times 64$ & 305.3(4) & 0.0102 &
      0.0509 & 0.635 \\
      a12m220 & 0.1184(10) & $32^3\times 64$ & 216.9(2) & 0.00507 &
      0.0507 & 0.628 \\
      a12m130 & 0.1191(7) & $48^3\times 64$ & 131.7(1) & 0.00184 &
      0.0507 & 0.628 \\
      \hline
      a09m310 & 0.0888(8) & $32^3\times 96$ & 312.7(6) & 0.0074 &
      0.037 & 0.440 \\
      a09m220 & 0.0872(7) & $48^3\times 96$ & 220.3(2) & 0.00363 &
      0.0363 & 0.430 \\
      \hline
      a06m310 & 0.0871(6) & $48^3\times 144$ & 319.3(5) & 0.0048 &
      0.024 & 0.286 \\
      \hline\hline
    \end{tabular}
  \end{center}
  \caption{Parameters of the MILC HISQ ensembles with $N_f=2+1+1$ 
    \cite{Bazavov:2012xda} used in our calculations.  The lattice
    spacing $a$ is set by the Sommer scale $r_1$ and  $N_s$ ($N_t$) is
    the lattice size in the spacial (temporal) direction.  $M_{\pi}$
    is the mass of Goldstone pions and  $am_l$, $am_s$ and $am_c$ are sea
    quark masses for the light (up and down), strange and charm quarks
    in lattice units, respectively.
    \label{tab:ensembles}}
\end{table}

The OK action $S_\text{OK}$ improves on the Fermilab formulation
of the Wilson clover action \cite{ ElKhadra:1996mp} by including
$\mathcal{O}(\lambda^2)$ and $\mathcal{O}(\lambda^3)$ improvement
terms in heavy quark effective theory (HQET) power counting.
\begin{align}
  S_{\text{OK}} & = a^{4} \sum_{x} \bar{\psi} (x) \left[
    \vphantom{\sum_{j\neq k} \left\{ i \Sigma_{k} B_{k} \,,\Delta_{j}
      \right\} } m_{0} + \gamma_{4} D_{4} \right. & \leftarrow
    \mathcal{O}(\lambda^0) \nonumber \\
    & \hphantom{\;=\ a^{4}\sum_{x}\bar{\psi}(x)} - \frac{1}{2} a
    \Delta_{4} + \zeta \bm{\gamma \cdot D} - \frac{1}{2} r_{s} \zeta a
    \Delta^{(3)} - \frac{1}{2} c_{B} a \zeta i \bm{\Sigma \cdot B} &
    \leftarrow \mathcal{O}(\lambda^1) \nonumber \\
    & \hphantom{\;=\ a^{4}\sum_{x}\bar{\psi}(x)} - \frac{1}{2} c_{E} a
    \zeta \bm{\alpha \cdot E} & \leftarrow \mathcal{O}(\lambda^2)
    \nonumber \\
    &\hphantom{=\ a^{4}\sum_{x}\bar{\psi}(x)}
    \begin{rcases}
      & + c_{1} a^{2} \sum_{k} \gamma_{k} D_{k} \Delta_{k} + c_{2}
      a^{2} \left\{ \bm{\gamma \cdot D},\Delta^{(3)} \right\} + c_{3}
      a^{2} \left\{ \bm{\gamma \cdot D} \, , i \bm{\Sigma \cdot B}
      \right\} \\
      & \left. + c_{EE} a^{2} \left\{ \gamma_{4} D_{4} \, , \bm{\alpha
        \cdot E} \right\} + c_{4} a^{3} \sum_{k} \Delta_{k}^{2} +
      c_{5} a^{3} \sum_{k} \sum_{j \neq k} \left\{ i \Sigma_{k} B_{k}
      \, , \Delta_{j} \right\} \right] \psi(x)
    \end{rcases} \,.
    & \leftarrow \mathcal{O}(\lambda^3)
\label{eq:ok-ac-1}
\end{align}
The definition of the operators in Eq.~\eqref{eq:ok-ac-1} can be found
in Ref.~\cite{Oktay:2008ex}.
The bare quark mass $m_0$ is related to the hopping parameter
$\kappa$ as follows,
\begin{align}
  a m_{0} = \frac{1}{2} \left( \frac{1}{\kappa} -
  \frac{1}{\kappa_{\text{crit}}} \right)\,.
  \label{eq:bare-m-1}
\end{align}
The non-perturbatively tuned hopping parameters for bottom and charm
quarks, $\kappa_b$, $\kappa_c$, and the critical hopping parameter
$\kappa_\text{crit}$~\cite{Bailey:2017xjk} for each measurement are
summarized in Table \ref{tab:parameters}.

In order to achieve a better overlap with the wave functions of the
$B_{(s)}$ and $D_{(s)}$ meson states, we apply the covariant Gaussian
smearing (CGS), $\left\{1 + \sigma^2\nabla^2/(4 N_{\text{GS}})
\right\}^{N_{\text{GS}}}$ to the point source and sink as in
Ref.~\cite{Yoon:2016dij}.
The CGS parameters $\{ \sigma, N_\text{GS} \}$ for each measurement
are given in Table \ref{tab:parameters}.
Here, we apply the CGS only to the heavy quark fields of the
pseudoscalar interpolating operators.
\section{Correlator and current improvement}
\label{sec:curim}
%
The meson-meson (MM) and meson-current (MC) 2-point correlators are
defined as follows \cite{Bazavov:2011aa},
\begin{align}
  C_{\text{MM}}(t) &= \sum_{\bf{x}} \left\langle
  \mathcal{O}^{\dagger}_P(t,{\bf x}) \mathcal{O}_P(0)
  \right\rangle 
  = \sum_{\alpha = 1}^{4} \sum_{\bf{x}} \left\langle
  \mathcal{O}^{\dagger}_\alpha(t,{\bf x}) \mathcal{O}_\alpha(0)
  \right\rangle \,, \\
  C_{\text{MC}}(t) &= \sum_{\bf{x}} \left\langle A^{4\dagger}(t,{\bf x})
  \mathcal{O}_P(0) \right\rangle
  = \sum_{\alpha = 1}^{4}
  \sum_{\bf{x}} \left\langle A^{4\dagger}_\alpha(t,{\bf x})
  \mathcal{O}_\alpha(0) \right\rangle\,,
  \label{eq:corr-2pt}
\end{align}
where the pseudoscalar heavy-light meson interpolating operator
$\mathcal{O}_\alpha(t,{\bf x})$ and the axial current operator
$A^4_{\alpha}(t,{\bf x})$ are
\begin{align}
  \mathcal{O}_\alpha(t,{\bf x}) &= \left[ \bar{\psi}(t,{\bf x})
    \gamma_5 \Omega(t,{\bf x}) \right]_\alpha \chi(t,{\bf x}) \,\\
  A^4_{\alpha}(t,{\bf x}) &= \left[ \bar{\Psi}(t,{\bf x}) \gamma^4
    \gamma_{5} \Omega(t,{\bf x}) \right]_{\alpha} \chi(t,{\bf x}) \,.
\end{align}
Here $\psi$ is the OK heavy quark field, $\chi$ is the HISQ 
light quark field, and 
\begin{align}
\Omega(t,{\bf x}) \equiv \gamma_{1}^{\;x_1} \gamma_{2}^{\;x_2}
\gamma_{3}^{\;x_3} \gamma_{4}^{\;t} \,,
\label{eq:omega-trans-1}
\end{align}
and the subscript $\alpha$ represents the taste degree of the
staggered light quarks.
The rotated heavy quark field $\Psi$ is introduced to improve the
axial current $A^4_{\alpha}$ up to $O(\lambda^3)$, the same
level as the OK action.
\begin{align}
  \Psi(t,{\bf x}) & = \Big( 1  & \leftarrow
  O(\lambda^{0}) \nonumber \\
  & \hphantom{=(\ \; } + d_{1} a \bm{\gamma \cdot D} & \leftarrow
  O(\lambda^{1}) \nonumber \\
  & \hphantom{=(\ \; } + d_{2} a^{2} \Delta^{\left(3\right)} + d_{B}
  a^{2} i \bm{\Sigma \cdot B} - d_{E} a^{2} \bm{\alpha \cdot
    E} & \leftarrow O(\lambda^{2}) \nonumber \\
  &\hphantom{=(\ \, }
  \begin{rcases}
    & + d_{rE} a^{3} \left\{ \bm{\gamma \cdot D}, \bm{\alpha \cdot E}
    \right\} - d_{3} a^{3} \sum_{i} \gamma_{i} D_{i} \Delta_{i} -
    d_{4} a^{3} \left\{ \bm{\gamma \cdot D} , \Delta^{\left(3\right)}
    \right\} \\
    & - d_{5} a^{3} \left\{ \bm{\gamma \cdot D} , i \bm{\Sigma \cdot
      B} \right\} + d_{EE} a^{3} \left\{ \gamma_{4} D_{4} , \bm{\alpha
      \cdot E} \right\} - d_{6} a^{3} \left[ \gamma_{4} D_{4} ,
      \Delta^{\left(3\right)} \right] \\
%
    & - d_{7} a^{3} \left[ \gamma_{4} D_{4} , i \bm{\Sigma
        \cdot B} \right] \Big) \psi (t,{\bf x} ) \,,
  \end{rcases}
  & \leftarrow O(\lambda^3)
  \label{eq:cur-imp-1}
\end{align}
where the improvement coefficients $d_i$ are given in Ref.~\cite{
  Bailey:2020uon}.

\begin{table}[!t]
  \begin{center}
    \begin{tabular}{ l || r | l l l | l | r }
      \hline\hline
      ensemble ID & $m_x / m_s$ & $\kappa_{\text{crit}}$ & $\kappa_c$
      & $\kappa_b$ & $\{ \sigma \,,\; N_{\text{GS}} \}$ &
      $N_{\text{cfg}} \times N_{\text{src}}$ \\
      \hline
      a12m310 & 1/5, 1 & 0.051211 & 0.048524 & 0.04102 &
      $\{ 1.5\,,\;5 \}$ & $1053 \times 3 $ \\
      a12m220 & 1/10, 1 & 0.051218 & 0.048613 & 0.04070 &
      $\{ 1.5\,,\;5 \}$ & $1000 \times 3 $ \\
      a12m130 & 1/27, 1 & 0.05119 & 0.048501 & 0.041343 &
      $\{ 1.5\,,\;5 \}$ & $499 \times 3 $ \\
      \hline
      a09m310 & 1/5, 1 & 0.05075 & 0.04894 & 0.0429 &
      $\{ 2.0\,,\;10 \}$ & $996 \times 3$ \\
      a09m220 & 1/10, 1 & 0.05077 & 0.04902 & 0.0431 &
      $\{ 2.0\,,\;10 \}$ & $1001 \times 3$ \\
      \hline
      a06m310 & 1/5, 1 & 0.050357 & 0.04924 & 0.0452 &
      $\{ 3.0\,,\;22 \}$ & $1017 \times 3$ \\
%
      \hline\hline
    \end{tabular}
  \end{center}
  \caption{The 2${}^{\rm nd}$ column gives the valence light quark masses
    $m_x$ and the following columns are the hopping parameters, CGS
    parameters and the number of measurements. $N_{\text{cfg}}$ represents
    the number of gauge configurations analyzed and $N_{\text{src}}$ is the number
    of sources used for measurement on each gauge configuration.
    \label{tab:parameters}}
\end{table}

\section{Correlator fit}
\label{sec:corfit}
\begin{figure}[t] 
  \subfigure[$B$ meson]{
    \label{fig:b-prior-1}
    \includegraphics[width=0.227\textwidth]
                    {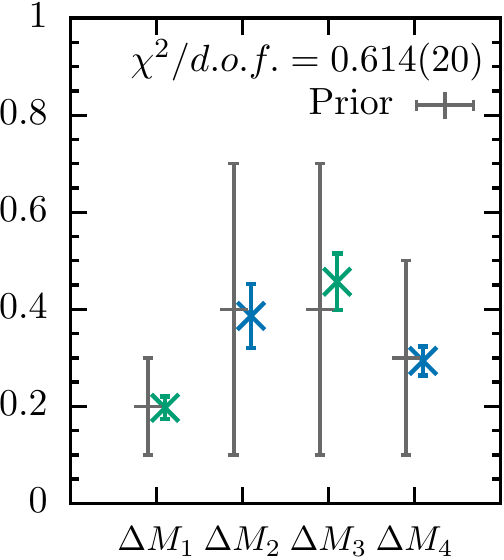}
  }
%
  \subfigure[$B_s$ meson]{
    \label{fig:b-prior-2}
    \includegraphics[width=0.227\textwidth]
                    {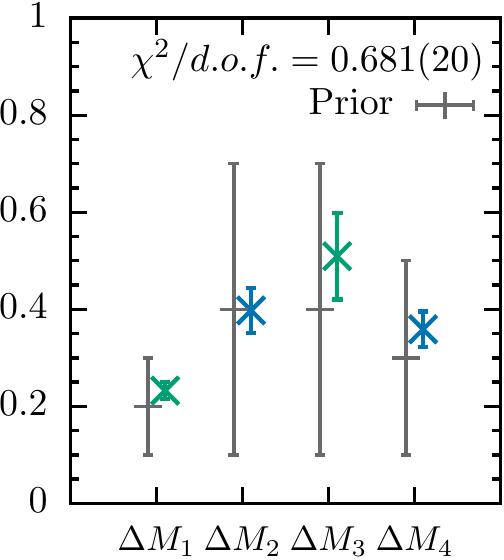}
  }
%
%
  \subfigure[$D$ meson]{
    \label{fig:d-prior-1}
    \includegraphics[width=0.227\textwidth]
                    {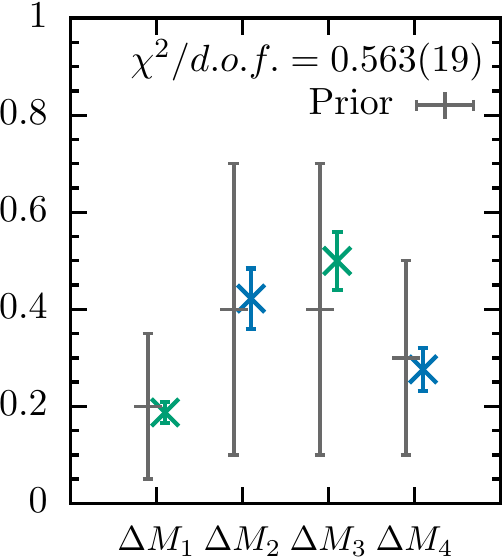}
  }
%
%
  \subfigure[$D_s$ meson]{
    \label{fig:d-prior-2}
    \includegraphics[width=0.227\textwidth]
                    {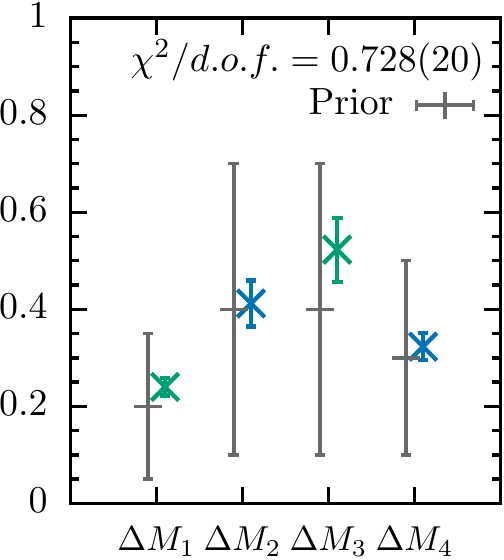}
  }
  \caption{ Fit results and Bayesian priors for $\Delta M_i$ from
    $a12m130$ ensembles.}
  \label{fig:prior-plots-1}
\end{figure}

%
We fit the 2-point correlation functions $C_\text{MM}(t)$ and
$C_\text{MC}(t)$ with three even time-parity and two odd time-parity
states and label it the 3+2-state fit.
The time parity is determined with respect to the shift operator in
the Euclidean time direction.
The fitting function is
\begin{align}
  C_{\text{Y}}(t) & = g_{\text{Y}}(t) \pm
  g_{\text{Y}}(T-t)\,, \qquad (\text{$+$ for MM, $-$ for MC)}
  \nonumber \\
  g_{\text{Y}}(t) & = A_{0}^{\text{Y}} e^{-M_0t}
  \left[ 1 + R_{2}^{\text{Y}}
    e^{-\Delta M_{2}t} + R_{4}^{\text{Y}}
    e^{-(\Delta M_{2} + \Delta M_{4})t}
  \vphantom{e^{-\Delta M_1^p t}} + \cdots \right. \nonumber \\
 & \hphantom{ = A_{0}^{\text{Y}}e^{-M_0t} [ } \left. \ -
       \left(-1\right)^t R_{1}^{\text{Y}} e^{-\Delta M_1 t} -
       \left(-1\right)^t R_{3}^{\text{Y}}
       e^{-(\Delta M_1 + \Delta M_3) t} + \cdots
       \right]
  \label{eq:3+2-fit}
\end{align}
where Y=MC or MM, $\Delta M_i \equiv M_i - M_{i-2}$, $M_{-1}\equiv
M_0$, and
\begin{align}
  A^\text{MM}_i &\equiv \frac{1}{2M_i}
  \langle 0 | \mathcal{O}_P | P_i \rangle
\langle P_i | \mathcal{O}_P | 0 \rangle\,,
\qquad
A^\text{MC}_i \equiv \frac{1}{2M_i}
\langle 0 | A^4 | P_i \rangle
  \langle P_i | \mathcal{O}_P | 0 \rangle \,,
\qquad
R^\text{Y}_i \equiv \frac{ A^\text{Y}_i }{ A^\text{Y}_0 }
\end{align}
Here, $P_i$ represents the $i$-th excited meson state and $P_0$ the ground state. For a brevity, the subscript ``0'' for
the ground state is dropped from now on.
We take the following steps to analyze the 2-point correlation
functions.
\begin{enumerate}
\item We fit the 2-point correlator, $C_\text{MM}(t)$, data using the
  3+2-state fit given in Eq.~\eqref{eq:3+2-fit} to
  extract the ground state pseudoscalar meson mass $M(\equiv M_0)$ and
  amplitude $A^\text{MM}(\equiv A^\text{MM}_0)$ and control the 
  excited states. We impose empirical Bayesian priors \cite{
    Yoon:2016jzj} on the excited state mass gaps $\Delta M_i$ and
  amplitude ratios $R^\text{MM}_i$ to stabilize the fit. (See
  Fig.~\ref{fig:prior-plots-1}).
\item We feed the results for $M_0$, and $\Delta M_i$ obtained
  in the previous step as inputs into the fit for $C_\text{MC}(t)$
  to extract $A^\text{MC} (\equiv A^\text{MC}_0)$ and the ratios
  $R^\text{MC}_i$.
  We use the same fit range and fit functional form as taken
  for $C_\text{MM}(t)$
\end{enumerate}

\begin{figure}[t] 
  \subfigure{
    \label{fig:meff-mm-1}
    \includegraphics[width=0.475\textwidth]
                    {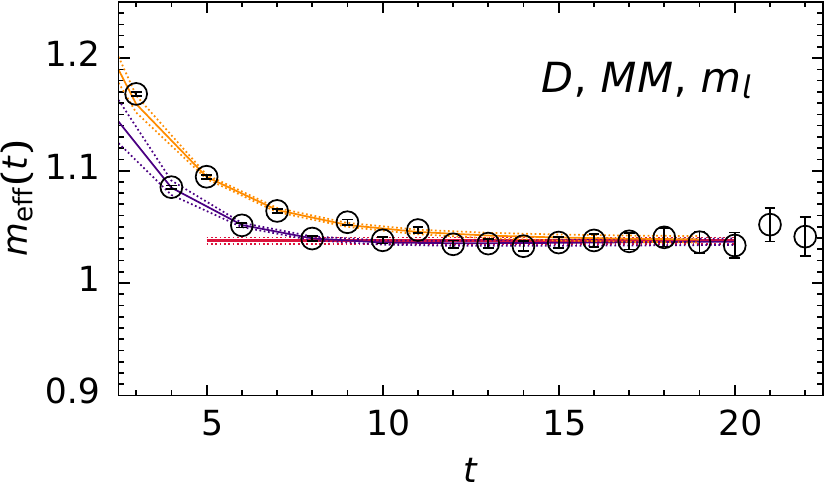}
  }
  \hfill
  \subfigure{
    \label{fig:meff-mc-1}
    \includegraphics[width=0.475\textwidth]
                    {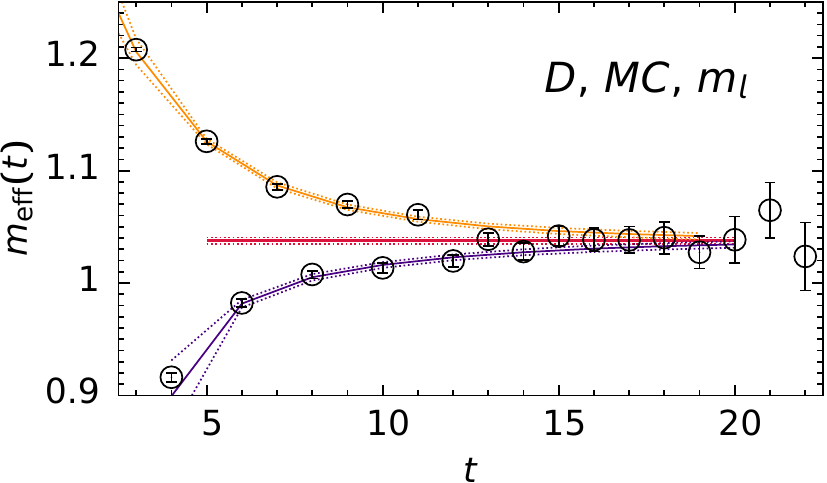}
  }
  \caption{ Effective mass plots for the $C_\text{MM}(t)$ and
    $C_\text{MC}(t)$ correlators of the $D$ meson on the $a12m130$
      ensemble with $m_l=m_s/27$.
    The orange (purple) curves connect 3+2-state fit results on the
    odd (even) time slices.
    The horizontal red line shows the ground state mass $M_0$
      within the fit range $[t_\text{min},t_\text{max}]=[5,20]$.
    The axial current operators are improved up to the $\lambda^3$
    order.
    \label{fig:meff-plots-1}}
\end{figure}
An example of the effective mass plot with 
\begin{align}
  m_{\text{eff}}^{Y}(t) &\equiv \frac{1}{2}
  \log\left[\frac{C_Y(t)}{C_Y(t+2)}\right] 
\end{align}
for  the 2-point correlators $C_\text{MM}(t)$ and
$C_\text{MC}(t)$ is shown in Fig.~\ref{fig:meff-plots-1} along with the fits to them.

\section{Results}
\label{sec:results}
%
The decay constant $f_P$ defined in Eq.~\eqref{eq:f} can be expressed
in terms of the ground state amplitudes $A^\text{MM}$ and
$A^\text{MC}$ as follows,
\begin{align}
   f_{P} = Z_{A^4}^{hl} \sqrt{\frac{2}{M_P}} \,
  \frac{A^\text{MC}}{\sqrt{A^\text{MM}}} \,,
\end{align}
where we take the meson mass $M_P=M_0$ of the ground state obtained
from the fits to $C_\text{MM}$.
The tree-level renormalization factor is given as
$Z_{A^4}^{hl,\text{tree}} = e^{m_{1}^h/2}$ where $m_1^{h} = \log(1 +
m_0^{h})$ is the rest mass and $m_0^h$ is the bare mass for the heavy
quark \cite{ElKhadra:1996mp}.
The perturbative and non-perturbative determination of $Z_{A^4}^{hl}$
is in progress.
In this work, we present the flavor $SU(3)$ breaking ratio of decay
constants:
\begin{align}
  f_{X_s} / f_{X} &= \frac{Z_{A^4}^{hs}}{Z_{A^4}^{hl}} \sqrt{
    \frac{M_{X}}{M_{X_s}} }
  \sqrt{\frac{A^\text{MM}}{A^\text{MM}_s}}
  \frac{A^\text{MC}_s}{A^\text{MC}}
  \cong \sqrt{
    \frac{M_{X}}{M_{X_s}} }
  \sqrt{\frac{A^\text{MM}}{A^\text{MM}_s}}
  \frac{A^\text{MC}_s}{A^\text{MC}}\,,
\end{align}
for $X_{(s)}= B_{(s)}$ and $D_{(s)}$ mesons.
Here, $A^\text{MM}_s$ and $A^\text{MC}_s$ are the ground state
amplitudes for the heavy-strange mesons.
In this ratio, we assume that the light quark mass $m_l$ dependence
of $Z_{A^4}^{hl}$ is negligible \cite{Bazavov:2011aa},  so the
ratio $Z_{A^4}^{hs} / Z_{A^4}^{hl} \cong 1$.
\begin{figure}[t!] 
  \subfigure[$f_{B_s} / f_B$]{ \label{fig:dec-b-2}
    \includegraphics[width=0.475\textwidth]{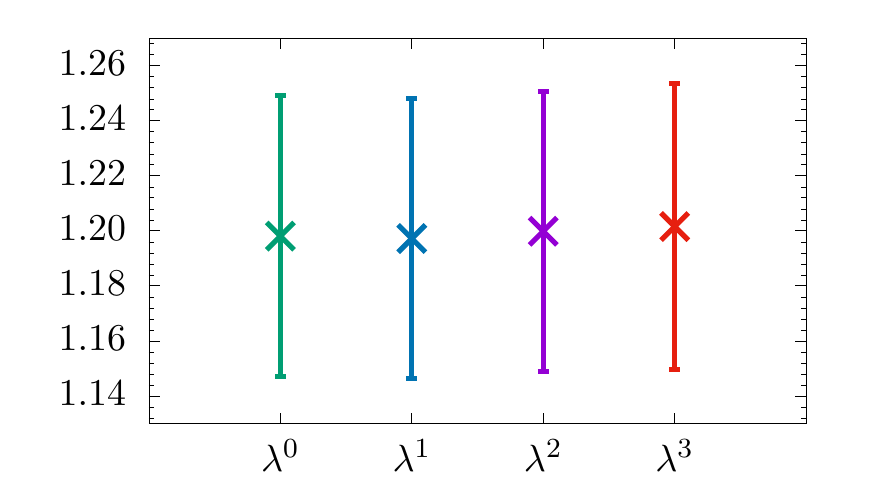} }
  \hfill
  \subfigure[$f_{D_s} / f_D$]{
    \label{fig:dec-d-2}
    \includegraphics[width=0.475\textwidth]{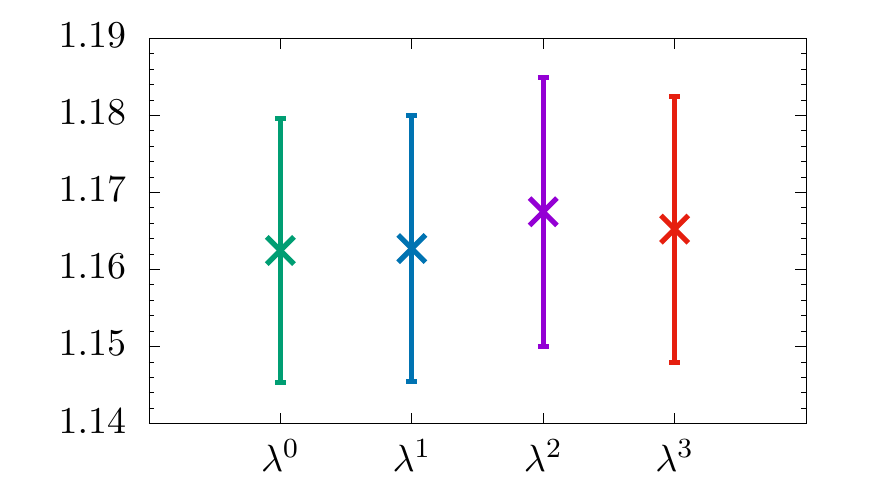} }
  \caption{ The ratios of decay constants, $f_{B_s} / f_B$ and
    $f_{D_s} / f_D$, on the a12m130 ensemble as a function of current
    improvement order in HQET power counting. }
  \label{fig:results-2}
\end{figure}
Fig.~\ref{fig:results-2} shows that in the ratios $f_{B_s} / f_{B}$ and $f_{D_s} / f_{D}$, 
the effect of the current improvement applied to the heavy quark field (as given in
Eq.~\eqref{eq:cur-imp-1}) cancels up to $O(\lambda^3)$. 

\begin{figure}[h!] 
  \subfigure[$f_{B_s}/f_B$]{ \label{fig:dec-b-1}
    \includegraphics[width=0.475\textwidth]{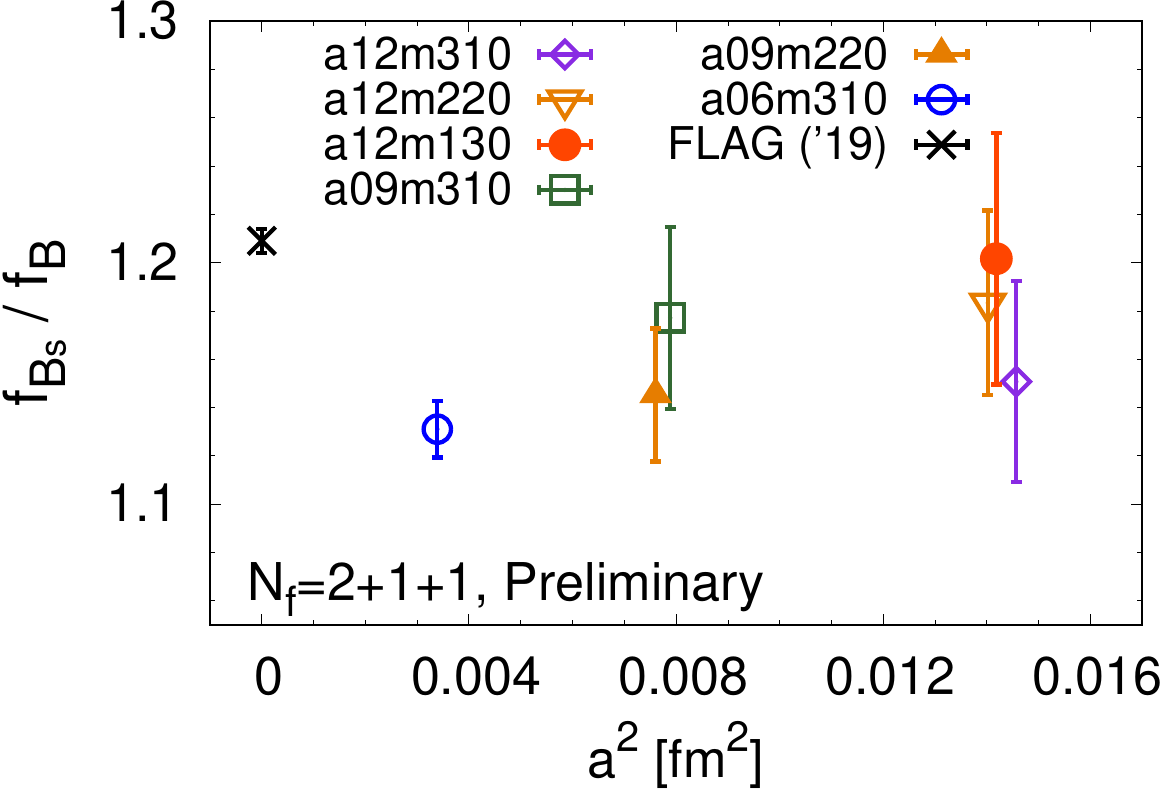} }
  \hfill
  \subfigure[$f_{D_s}/f_D$]{
    \label{fig:dec-d-1}
    \includegraphics[width=0.475\textwidth]{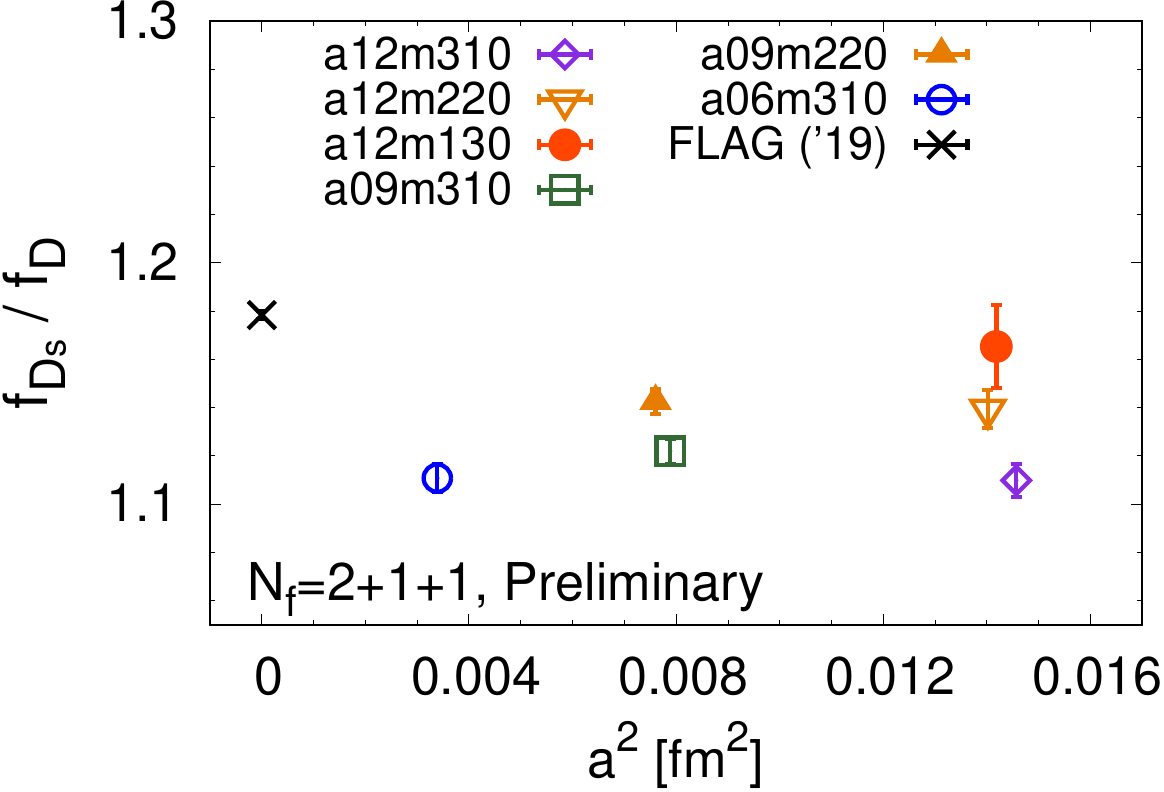} }
  \caption{The ratios of decay constants, $f_{B_s} / f_B$ and $f_{D_s}
    / f_D$ on six ensembles.
    The errors are purely statistical.
    The FLAG 2019 \cite{ Aoki:2019cca} results are given for the
    physical value.}
  \label{fig:results-1}
\end{figure}

In Fig.~\ref{fig:results-1}, we present preliminary results on
$f_{B_s} / f_{B}$ and $f_{D_s} / f_{D}$ calculated on 6 different HISQ
ensembles, and compare them with the continuum limit value given in the FLAG
2019 review~\cite{Aoki:2019cca}.
The statistical errors in $f_{B_s} / f_{B}$
(Fig.~\ref{fig:results-1}\;\subref{fig:dec-b-1}) are much larger than in
$f_{D_s}/f_D$ (Fig.~\ref{fig:results-1}\;\subref{fig:dec-d-1}).
As a result, discerning a chiral or discretization effect in $f_{B_s}/f_B$ is not clear, 
other than to note that the result from the physical ensemble $a12m130$ is
consistent with the FLAG 2019 value.
The results for $f_{D_s}/f_{D}$ show no significant discretization
effect on the three lattices with $M_\pi \approx 310\,\MeV$ and the
two with $M_\pi \approx 220\,\MeV$. On the other hand, there is a
shift upwards towards the FLAG result as $M_\pi$ is lowered towards
the physical value.
Presuming that the OK action has significantly eliminated the heavy
quark discretization error even on the coarsest lattice spacing
$a\approx 0.12\,\fm$~\cite{Bailey:2017nzm}, the leading effect to
quantify is the pion mass dependence. For the $a \approx 0.12\,\fm$
data, the current trend is anchored by the physical ensemble with a
value close to the FLAG 2019 result.
In near future, we plan to add measurements on more ensembles to check for discretization
effects and on more physical pion mass ensembles to improve 
the chiral-continuum extrapolation.
%

\acknowledgments
\label{sec:ackn}
%
We thank the MILC collaboration for sharing the HISQ ensembles
with us.
Computations for this work were carried out in part on (i) facilities
of the USQCD collaboration, which are funded by the Office of Science
of the U.S. Department of Energy, (ii) the Nurion supercomputer at
KISTI and (iii) the DAVID GPU clusters at Seoul National University.
The research of W. Lee is supported by the Mid-Career Research Program
(Grant No.~NRF-2019R1A2C2085685) of the NRF grant funded by the Korean
government (MOE).
This work was supported by Seoul National University Research Grant in
2019.
W.~Lee would like to acknowledge the support from the KISTI
supercomputing center through the strategic support program for the
supercomputing application research (No.~KSC-2017-G2-0009).
T.~Bhattacharya and R.~Gupta were partly supported by the
U.S. Department of Energy, Office of Science, Office of High Energy
Physics under Contract No. DE-AC52-06NA25396.
S.~Park, T.~Bhattacharya, R.~Gupta and Y.-C.~Jang were partly
supported by the LANL LDRD program.
Y.-C.~Jang is partly supported by U.S.~Department of Energy under
Contract No.~DE-SC0012704.
%
\bibliography{ref}

\end{document}

%% file: macro.tex
\newcommand{\MeV}{\mathop{\rm MeV}\nolimits}
\newcommand{\fm}{\,\mathrm{fm}}

\newenvironment{rcases} {\left.\begin{aligned}}
               {\end{aligned}\right\rbrace}

